\documentclass[12pt]{iopart}

\usepackage{iopams}  
\usepackage{graphicx}

\begin{document}

\title[Theoretical confirmation of Feynman's hypothesis]{Theoretical
confirmation of Feynman's hypothesis on the creation of circular vortices
in Bose--Einstein condensates: III}

\author {E Infeld, A Senatorski and A A Skorupski}
\address{Department of Theoretical Physics, So{\l}tan Institute for Nuclear
Studies, Ho\.za 69, 00--681 Warsaw,
Poland}

\eads{\mailto{einfeld@fuw.edu.pl}, \mailto{senator@fuw.edu.pl}, \mailto{askor@fuw.edu.pl}}

\begin{abstract}
In two preceding papers (Infeld and Senatorski 2003 \textit{J. Phys.: Condens.
Matter} \textbf{15} 5865, Senatorski and Infeld 2004 \textit{\it J. Phys.: Condens. Matter}
\textbf{16} 6589) the authors confirmed Feynman's hypothesis on how circular vortices can be
created from oppositely polarized pairs of linear vortices (first paper), and then
gave examples of the creation of several \textit{different} circular vortices
from one linear pair (second paper). Here in part III, we give two classes
of examples of how the vortices can interact. The first
confirms the intuition that the reconnection processes which join two interacting
vortex lines into one and thus would increase the degree of entanglement of the vortex
system, practically do not occur. The second shows that new circular
vortices can also be created from pairs of oppositely polarized \textit{coaxial circular}
vortices. This seems to contradict the results for such pairs given in
Koplik and Levine 1996 \textit{Phys. Rev. Lett.} \textbf{76} 4745.
\end{abstract}

\submitto{\JPCM}


\section{Introduction}

The discovery of superfluidity in helium II ($^4$He and $^3$He) has
aroused interest in boson liquids. This was recently renewed and intensified due
to the experimental obtaining of Bose--Einstein condensates (BECs) in alkali metals.
Especially interesting is the formation and time evolution of both curved and straight
vortices which appear in these media \cite{Bew,Paol,Schwar}.

In two preceding papers
\cite{FEH1,FEH2}, referred to as part I and part II, the time evolution of a pair
of oppositely polarized linear vortices was examined.
In part I, it was demonstrated that if such a pair is appropriately perturbed,
a persistent set of
identical circular vortices can be created. This was in full agreement with
Feynman's hypothesis \cite{FEYNM} mentioned in the title. In part II, examples were given
for the creation of several \textit{different} circular vortices from one linear
pair. In both cases, the BEC was examined within the commonly used Gross--Pitaevskii model
\cite{GE,PL}. The authors checked that all circular vortices created during the
evolution satisfied known relations between their velocities and radii \cite{JRG}.

Here, in part III, we examine the interaction of circular vortices with either
linear ones, or else with other circular vortices of different radii, again within the
Gross--Pitaevskii model. The results are compared with those of other authors
\cite{KWS85}--\cite{Lead}.

\section{Basic equations and initial conditions}

In the present paper, as in parts I and II, we describe a one-component BEC
by a single particle wavefunction of $N$ bosons of mass $m$ that obeys the nonlinear
Schr\"odinger (NLS) type equation as formulated by Gross \cite{GE} and Pitaevskii \cite{PL},
\begin{equation}\label{nls}
\rmi \hbar \frac{\partial \psi}{\partial t} = - \frac{\hbar^2}{2m}
\nabla^2 \psi + W_0 \psi |\psi|^2,
\end{equation}
where positive $W_0$ is the strength of the assumed $\delta$-function repulsive potential
between bosons. Both linear pair and circular stationary vortex solutions are known, of
which only the last ones were shown to be stable (see part I for references). Similarly,
and for the same reasons as in parts I and II, we introduce dimensionless variables
defined by the transformation
\begin{equation}\label{trans}
\fl\psi \to \sqrt{\rho_0}\,\rme^{-\rmi \mu t/\hbar} \psi,
\qquad \mathbf{x} \to \frac{\hbar}{\sqrt{2 \mu m}}\mathbf{x},
\qquad t \to \frac{\hbar}{2 \mu } t,
\qquad \mu = W_0 \rho_0,
\end{equation}
where $\mu$ is the chemical potential and $\rho_0$ is the background mass density. Now, we obtain
from (\ref{trans}) and (\ref{nls})
\begin{equation}\label{nlstr}
2 \rmi \frac{\partial \psi}{\partial t} = - \nabla^2 \psi -
\psi (1 - |\psi|^2).
\end{equation}

An initial condition used in parts I and II, was
\begin{equation}\label{incond}
\psi (t = 0) = \frac{r_1 r_2}{\sqrt{r_1^2 + b^2} \sqrt{r_2^2 + b^2}}
\, \rme^{\rmi(\theta_1 + \theta_2)},
\end{equation}
where ($y$ here has been interchanged with $z$)
\begin{equation}\label{incona}
r_1^2 = (1 - 2 U^2)(x + a)^2 + y^2, \qquad
r_2^2 = (1 - 2 U^2)(x - a)^2 + y^2,
\end{equation}
\begin{equation}\label{inconf}
\tan \theta_{1,2} = \frac{y}{\sqrt{1 - 2 U^2}(a \pm x)},
\end{equation}
which is a model representing two oppositely polarized rectilinear vortices parallel to
the $xz$ plane and moving along the $y$ axis. The constant $b$ was chosen such that the
velocity of the two rectilinear vortices (along $y$) in the simulation would agree with
$U$, from the theory, see \cite{JRG}, table 2. In spite
of its simplicity, this model satisfies several criteria. The variable $x$ is scaled
correctly, the wavefunction is zero at the centres of the two vortices, and far field
behaviour is approximately correct (of the order of $1+ O(1/r^2), r = \sqrt{x^2 + y^2}$).
(We have doubts about the formula of Jones and Roberts, see Appendix.)
When the vortices are well separated, Fetters formula \cite{Fett,Infsk} is recovered
($b^2 \to 4$, and $U \to 0$ as $a \to \infty$).

In this paper, an initial condition for a circular vortex will be used. It will be described
by formulae of similar form to (\ref{incond})--(\ref{inconf}), but with $x$ and $y$
replaced by
\begin{equation}\label{inconb}
 x \to {\rm sgn} (x - x_0) \sqrt{(x - x_0)^2 + (z - z_0)^2}, \qquad y \to y - y_0.
\end{equation}
The vortex in question will move along the $y$ axis with
velocity close to $U$ taken from table 1 in \cite{JRG}, and the vortex lines
($\psi = 0$) at $t = 0$ will intersect the plane $z = z_0$ at two points: $x = x_0
\mp a,\: y = y_0$.

It is clear that the accuracy of such a model will increase as we increase
the radius $a$ of the circular vortex. However, as we observed in all considered cases, even
for not very large values of $a$, the oscillations of the respective circular vortices during
their motion were negligible.

The nonlinear Schr\"odinger equation in three dimensions (\ref{nlstr}) was numerically solved by
using a discrete fast Fourier transform in $x$, $y$, and $z$ to calculate the space derivatives
(pseudospectral algorithm), along with the leapfrog timestep. Calculations were performed in
a box: $0 \leq x \leq 2L_x$, $0 \leq y \leq 2L_y$, $0 \leq z \leq 2L_z$, with the number of
mesh points $N_x = N_z = 24$--$72$, and $N_y = 96$--$192$. Periodic boundary conditions were
assumed, and the timestep was determined from the numerical stability condition. The details of
our calculations are described in \cite{numalg}.

\section{Collisions of circular and almost linear vortices}

Figures 1 and 2 present collisions of a circular vortex parallel to the $xz$ plane
(of radius $r$) with an arc of another such vortex of much larger radius $R$ ($\gg r$).
This arc, along with its periodic continuations to neighbouring periodicity boxes, models a
linear vortex. (We have evidence that in this model, small disturbances of the initial condition
for linear vortex, also involving small discontinuities in derivatives at the boundaries, only
introduce small oscillations but do not change the main features of the evolution.)
As the advection velocity (along the $y$ axis) of a
circular vortex is a decreasing function of its radius (see e.g. figure 2 in Part I),
the circular vortex of radius $r$ will be much faster than the ``linear'' one which, in
the situation shown in figures 1 and 2, will imply their collision.

In figure 1, the $x$ components of the polarization vectors for the linear vortex and the
neighbouring part of the circular one during collision have opposite signs, which locally
resembles the situation analysed in Parts I and II,
i.e. a pair of oppositely polarized rectilinear vortices. In figure 2, the pertinent
signs are the same. Nevertheless, strange as it may seem, the result of the collision
in both cases is topologically the same, i.e. a (distorted) circular vortex and a linear vortex.
This is in contradiction to the description of the two above mentioned collisions given
by Schwarz, see the second and third situation presented in figure 16 of \cite{KWS85}. In both
these situations (in the Schwarz description) the reconnection occurs only at one of the crossing
points and the result of
the collision is topologically different from the initial state (one linear vortex only in the
final state). It should be mentioned that different behaviour at the two crossing points
(occurrence and lack of reconnection) is in fact also in contradiction to the results of Koplik
and Levine \cite{K&L93} who examined the evolution (within the GP model) of two neighbouring
rectilinear vortices for various angles between their vorticity vectors. Reconnection was
demonstrated for angles between $90^{\circ}$ and $180^{\circ}$, and no reconnection for
$45^{\circ}$. Thus, in the case of two rectilinear vortices, the angle $45^{\circ}$ implied the
same behaviour as that for $0^{\circ}$. For the collisions shown in our figures 1 and 2, the
angles between vorticity vectors at two intersection points are the same and one should not
expect different behaviour there. Another question, however, is the applicability of the results
of Koplik and Levine \cite{K&L93} to the collisions shown in figures 1 and 2, where one can only
locally think in terms of rectilinear vortices during collision.

\begin{figure}[t!]
\centering\includegraphics[scale=.65]{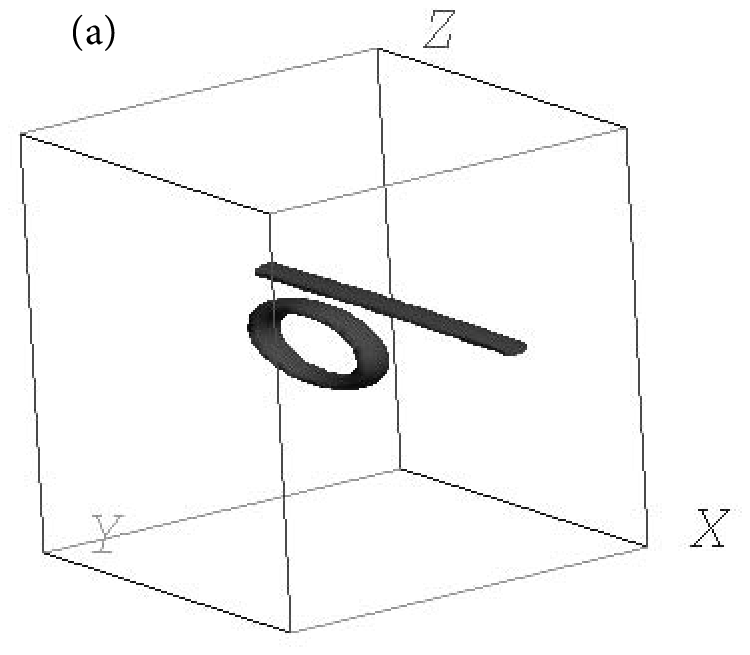}%
\quad\includegraphics[scale=.65]{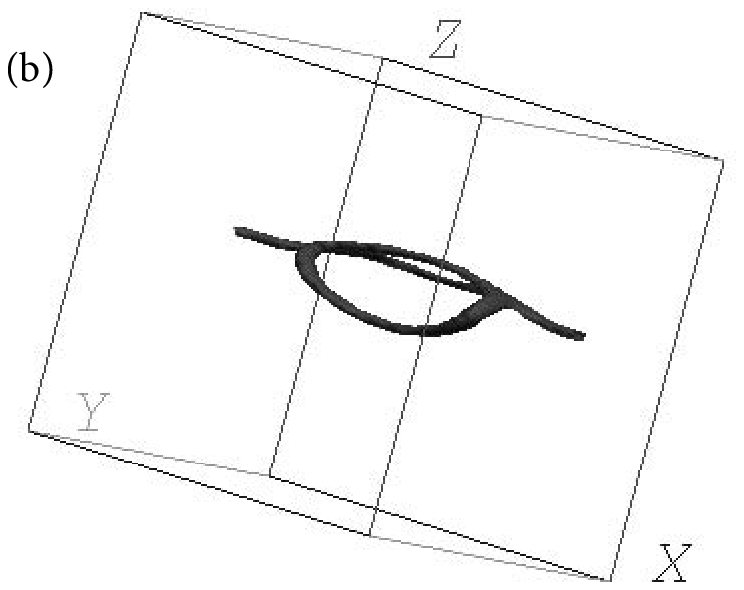}%
\quad\includegraphics[scale=.65]{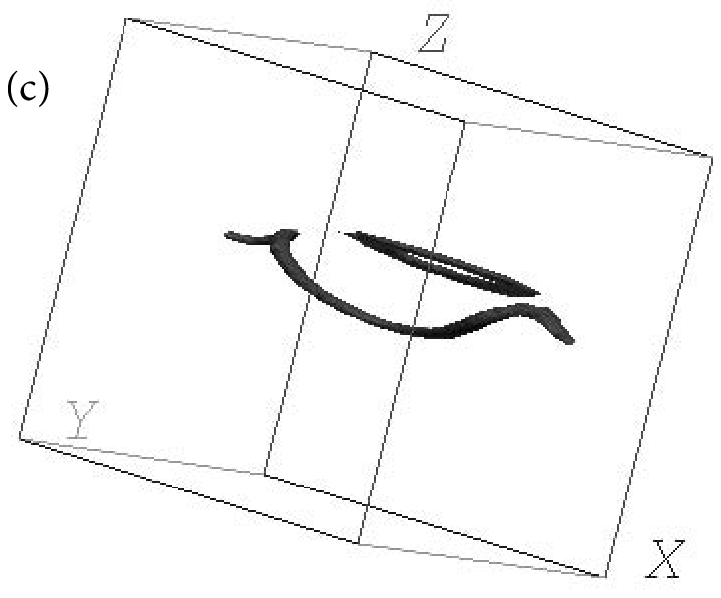}%
\hspace*{\fill}\\
\caption{Collision of a circular vortex parallel to the $xz$ plane
(of radius $r = 8$) with an arc of another such vortex, but of much larger radius $R = 400$,
such that $R/r = 50$. The distance between the ``rectilinear'' vortex and the axis of the
circular one is 4. Consecutive times are: 0, 5$\tau$, 7$\tau$, $\tau = 5.26$.
Polarizations of the nearest segments of the approaching vortices are opposed.}
\end{figure}
\begin{figure}
\centering\includegraphics[scale=.65]{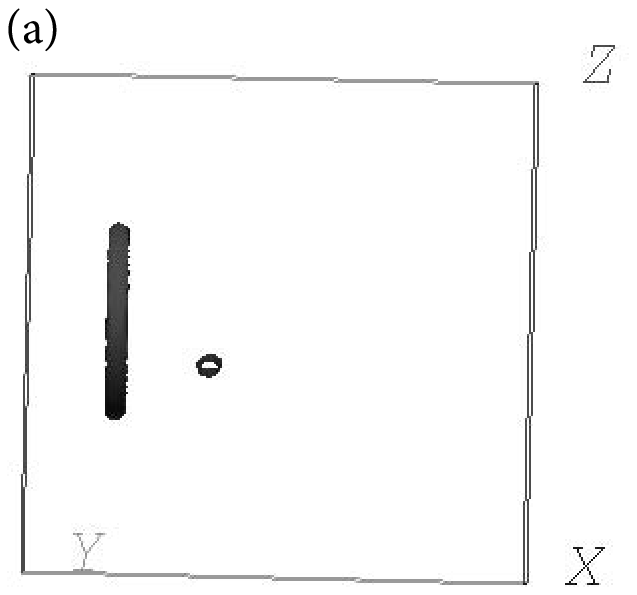}%
\quad\includegraphics[scale=.65]{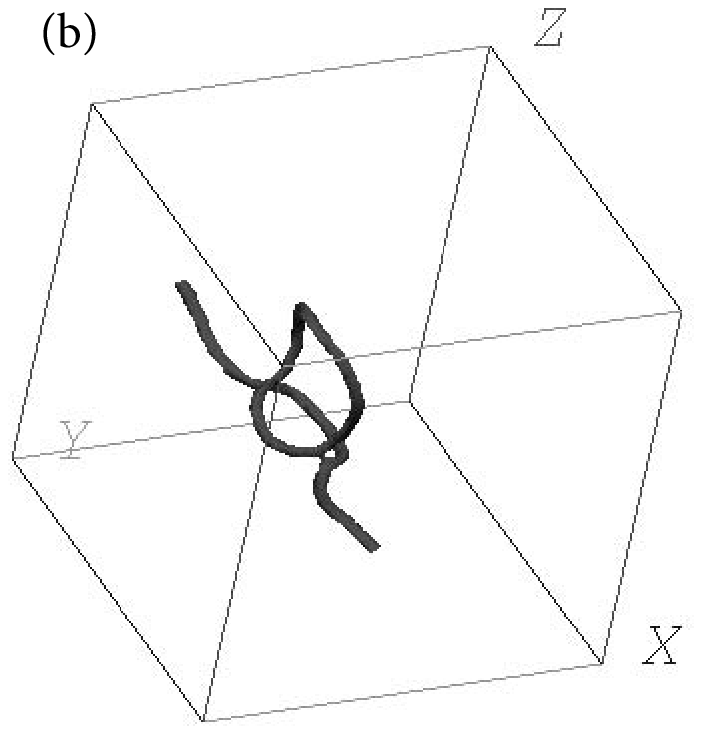}%
\quad\includegraphics[scale=.65]{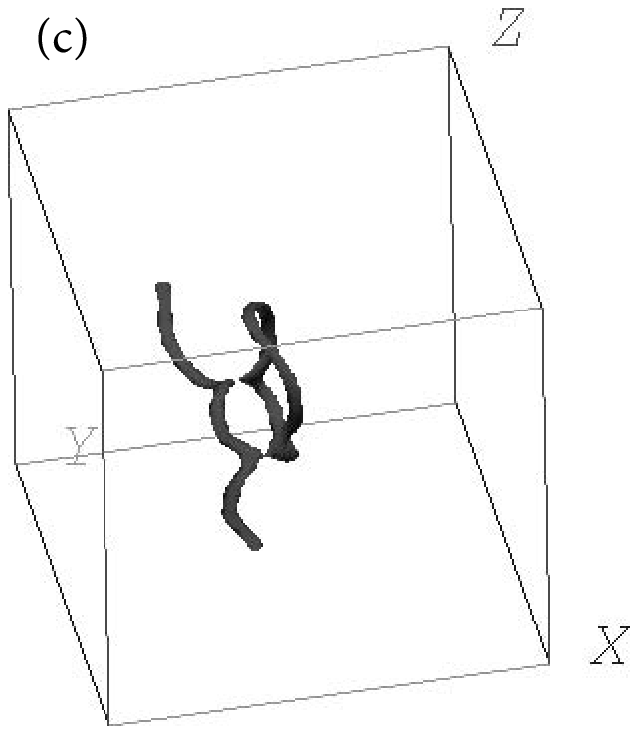}%
\quad\includegraphics[scale=.65]{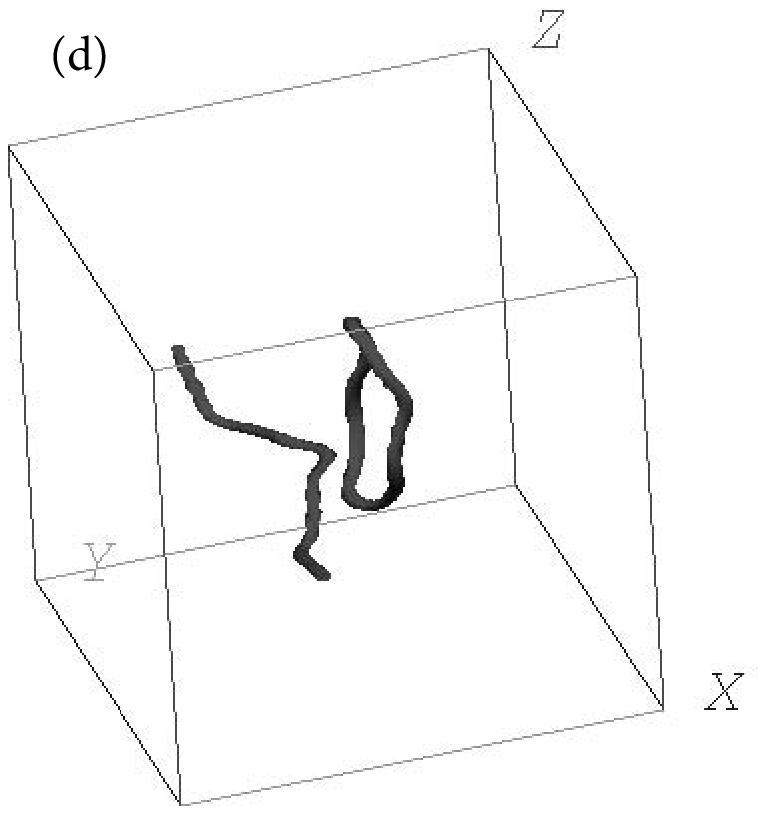}%
\hspace*{\fill}\\
\caption{Collision similar to that shown in Fig. 1 but where the polarizations
of the nearest segments of the colliding vortices agree. The radii are $r = 18$, and
$R = 900$ ($R/r = 50$), and the distance between the ``rectilinear'' vortex and the axis of the
circular one is 9. Consecutive times are: $0$, $14 \tau$, $18 \tau$,
$24 \tau$, $\tau = 6.11$. The first view is along the $x$ axis.}
\end{figure}

Here we pause for some comments. Firstly, we repeated
the calculation but for the segment of the large circular vortex with opposite
curvature. No difference between the results in both cases was observed. Secondly, we performed
a calculation for a similar initial vortex configuration, but with both colliding circular vortices
having comparable radii. We obtained essentially the same results (i.e. topologically, both
results could be treated as identical). So the exact values of the radii of the colliding vortices
are not important, if only the radius of one of the vortices is finite. Even more, a similar
scenario of the collision of two circular oppositely polarized vortices is observed if the vortices
have identical radii and their symmetry axes are parallel to each other but do not coincide. Again
two reconnections at the crossing points and two distorted circular vortices after collision are
obtained, see \cite{Lead}. Note also that the initial conditions presented in figures 1 and 2 are
symmetric with respect to the plane $x = L_x$ passing through both centres of colliding rings. This
implies that this symmetry lasts during the evolution. This was the case in our results even if it
is not obvious from figures 1 and 2.

As we have mentioned earlier, collisions as considered here were also investigated by K W Schwarz
\cite {KWS85} who performed the
first important step towards a formal description of the dynamics of vortices in BECs \cite%
{KWS85, KWS88}. Schwarz was able to explain many features of the vortex dynamics in these media,
including the interaction with walls. In order to describe some simple observed effects, such as
``avoiding collision'' between parallel vortices and ``attraction'' by antiparallel ones,
Schwarz introduced explicitly an ``artificial force'' acting between cores of neighbouring
vortex lines. This made it possible to reach many detailed results, but also introduced the risk of
some of these results being questionable.

The disagreement with our results has an important consequence. If  in the course of the collision
in question only one reconnection point appears, the two vortex lines join together into one
vortex line and the degree of entanglement of the system of these lines increases. Our result, on the
contrary, states that such coplanar collisions cannot increase the degree of entanglement and such a
quick generation of vortex entanglement as showed Schwarz in his figure 4 in \cite{KWS88} cannot come
about in this fashion.
\begin{figure}[t]
\centering\includegraphics[scale=.65]{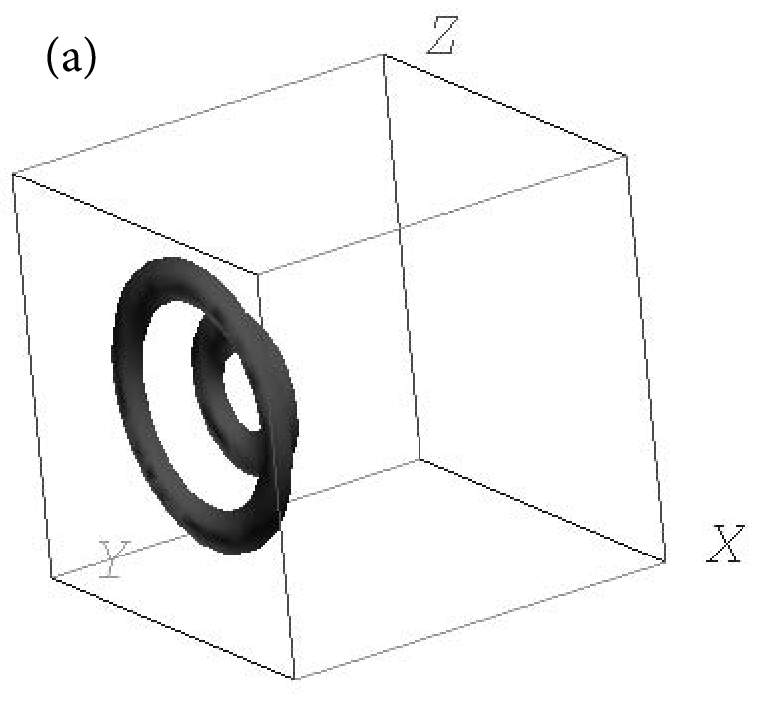}%
\quad\includegraphics[scale=.65]{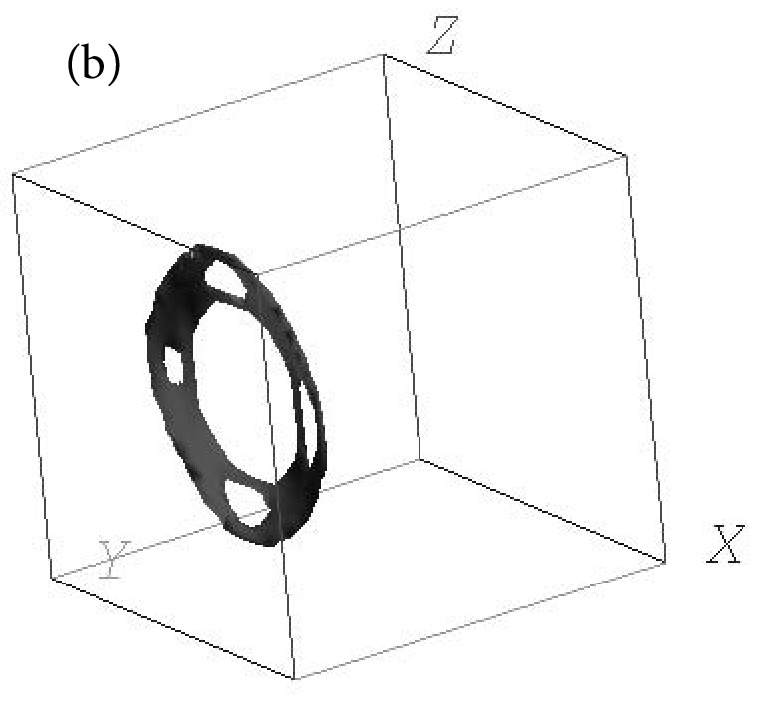}%
\quad\includegraphics[scale=.65]{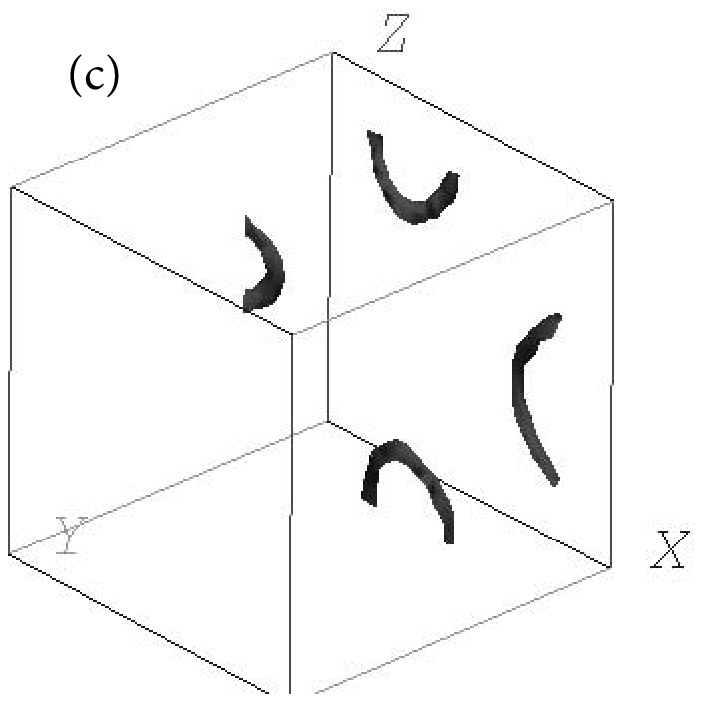}%
\hspace*{\fill}\\
\caption{Coaxial collision of two oppositely polarized circular vortices parallel to the $xz$
plane with comparable radii, $R = 8$, $r = 4$ ($R/r = 2$). The periodicity box has transverse
dimensions ($2L_x = 2L_z \equiv 2L$) comparable to the larger diameter $2R$ ($2L = 28$, i.e.
$L/R = 1.75$). Consecutive times are: $0$, $4 \tau$, $20 \tau$, $\tau = 5.23$.}
\end{figure}

\section{Collisions of coaxial oppositely polarized circular vortices}
In parts I and II the authors gave several examples of how circular vortices can be created from
pairs of oppositely polarized rectilinear vortices perturbed sinusoidally in space. Here we will
show that similar circular vortices can be created, if a circular vortex passes by another,
oppositely polarized coaxial circular vortex of somewhat larger radius. This seems to contradict
what Koplik and Levine say when considering a coaxial collision of oppositely polarized circular
vortices of different radii, (\cite{K&L96}, p. 4746):

``Two cases occur if two rings of \textit{different} sizes approach on axis: if the ratio of radii is
large, the rings simply leapfrog each other. If, however, the radii are comparable, one again sees
annihilation similar to the equal-sized case, except that the partially overlapping rings continue
to translate during the merger stage preceding annihilation.''

Different behaviour, resulting in reconnections and formation of circular vortices, should be possible
if the the collision of the circular rings is somehow perturbed. In the case of rectilinear vortices,
this perturbation was introduced externally. In the case of circular vortices, it could result from
interaction (in the sense of Schwarz \cite{KWS85, KWS88}) of the colliding vortices with other
circular vortices close by during the collision. Such situations are very probable in real
condensates. In this paper, they have been modelled in the simplest possible way, i.e. by choosing
the dimensions of the periodicity box in the plane parallel to the colliding vortices ($xz$ plane)
to be comparable to the diameters of the colliding vortices. With this
choice, the perturbation is due to the interaction of the colliding vortices  with their
neighbouring periodic images parallel to the $xz$ plane. There are four closest neighbours in
the directions of the $x$ and $z$ axis, four further ones in the directions rotated by $45^{\circ}$ along
the axis of the colliding vortices, still more along the $x$ and $z$ axis, etc. This evidently
suggests that the number of circular vortices after collision should be a multiple of four. This
prediction was confirmed by the results of our calculations as shown in figures 3--7.

\begin{figure}
\centering\includegraphics[scale=.6]{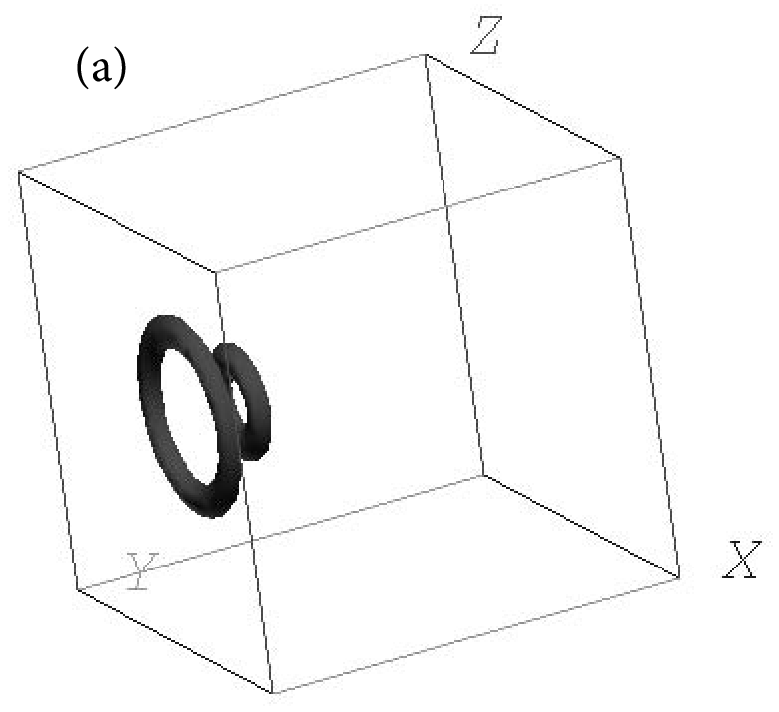}%
\quad\includegraphics[scale=.6]{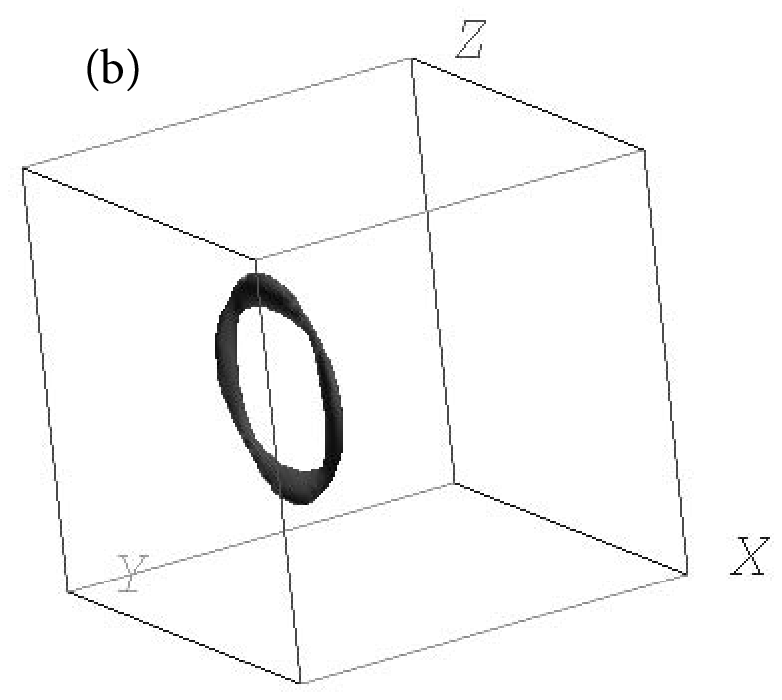}%
\quad\includegraphics[scale=.6]{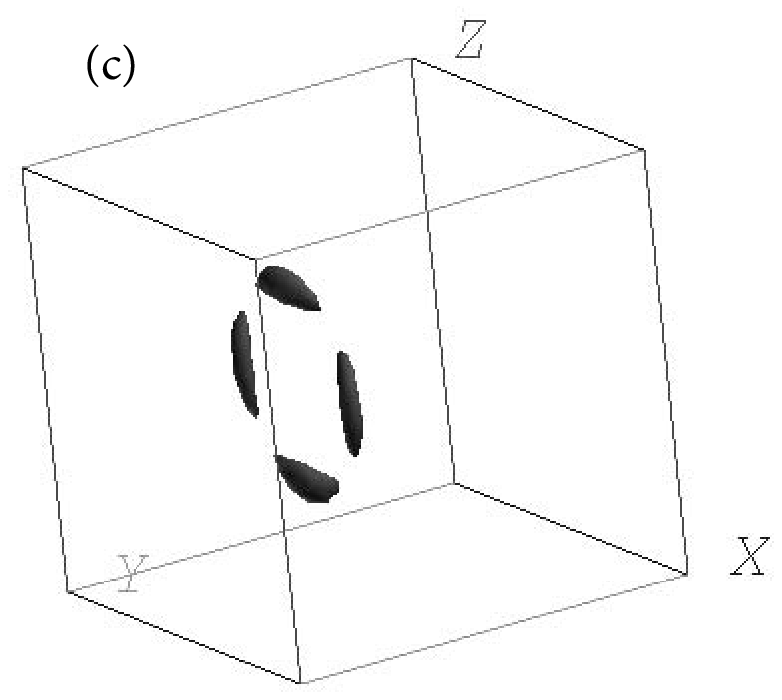}%
\quad\includegraphics[scale=.6]{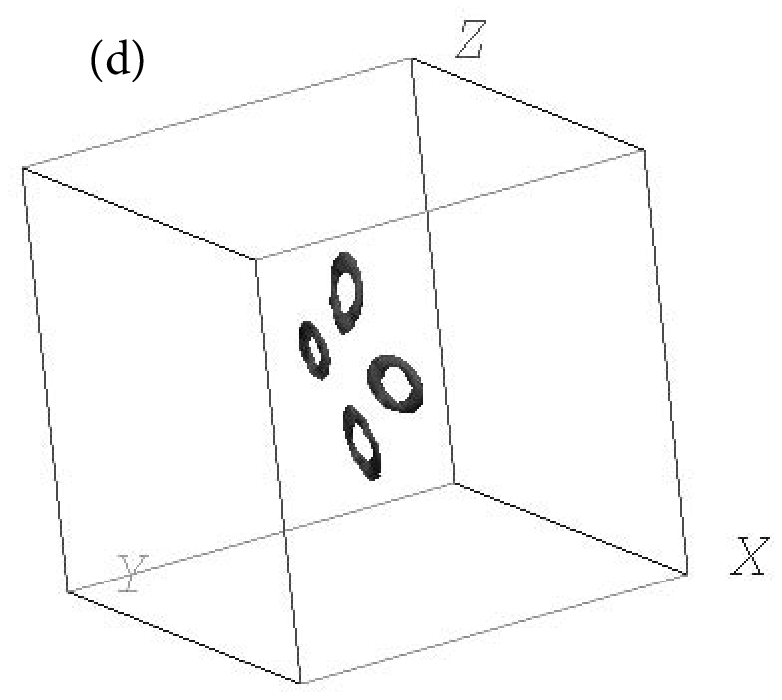}%
\hspace*{\fill}\\
\caption{Coaxial collision of two circular vortices identical with those shown in Fig. 3
($R/r = 2$) but with somewhat larger periodicity box ($2L = 42$, i.e. $L/R = 2.625$). The formation
of four smaller circular vortices after collision can be seen. Consecutive times are: $0$, $6 \tau$,
$7 \tau$, $11 \tau$, $\tau = 6.75$.}
\end{figure}
\begin{figure}
\centering\includegraphics[scale=.65]{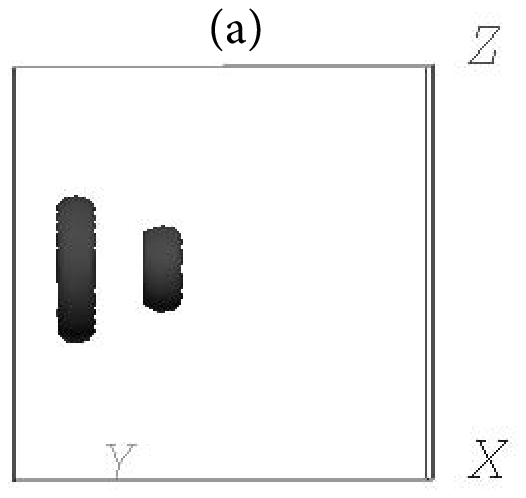}%
\quad\includegraphics[scale=.65]{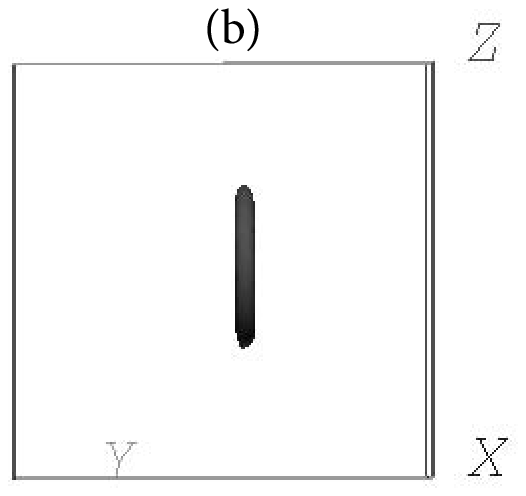}%
\quad\includegraphics[scale=.65]{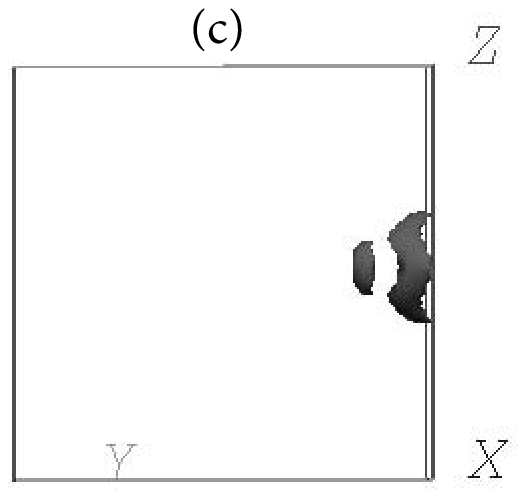}%
\hspace*{\fill}\\
\caption{Coaxial collision of two circular vortices identical with those shown in Fig. 3
($R/r = 2$) but with a relatively large periodicity box ($2L = 55$, i.e. $L/R \simeq 3.44$).
No smaller circular vortices after collision can be seen.
Consecutive times are: $0$, $6 \tau$, $11 \tau$, $\tau = 6.29$.}
\end{figure}
\begin{figure}

\centering\includegraphics[scale=.65]{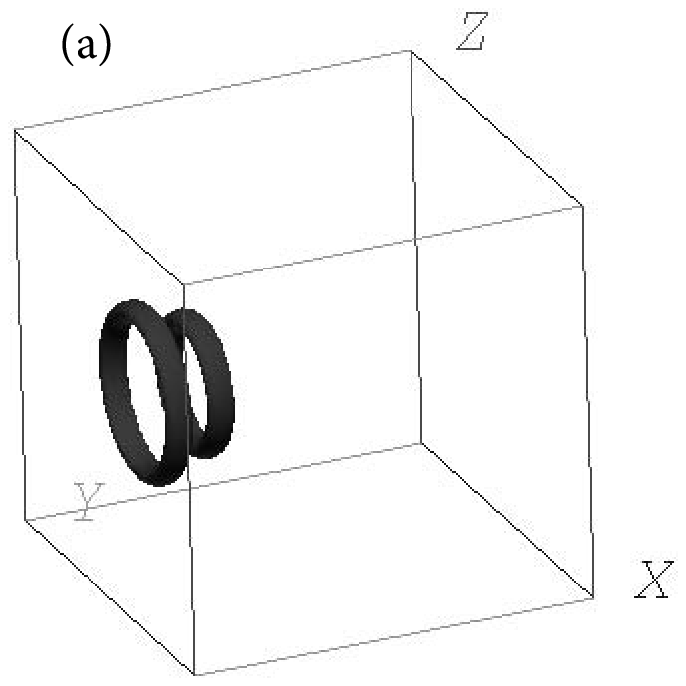}%
\quad\includegraphics[scale=.65]{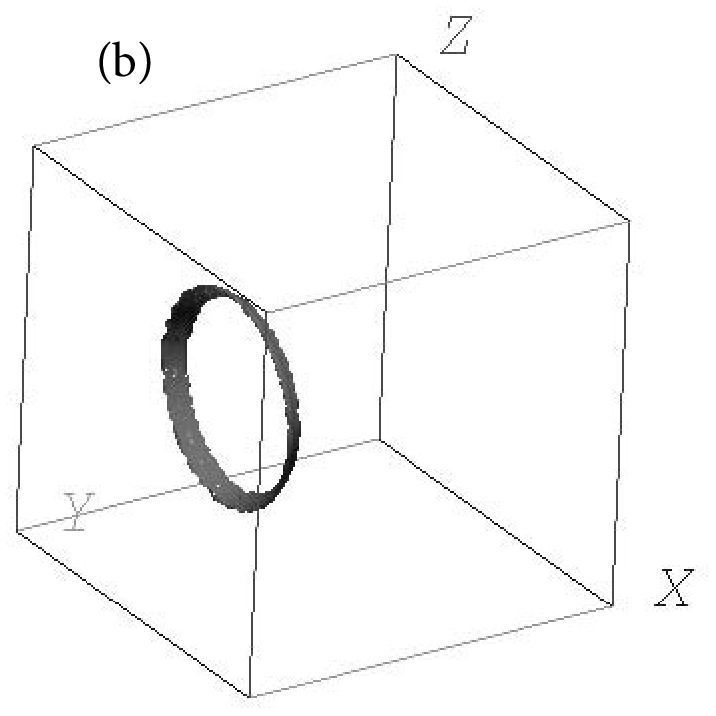}%
\quad\includegraphics[scale=.65]{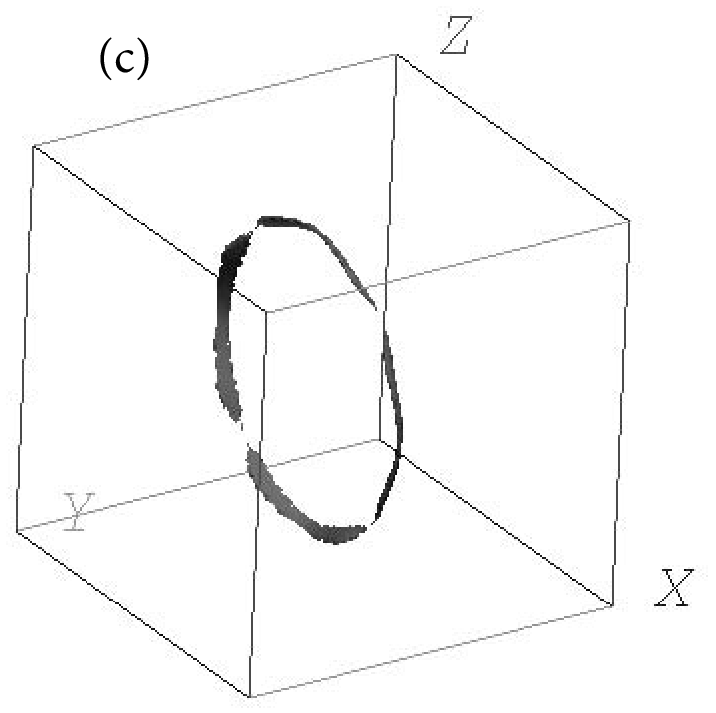}%
\quad\includegraphics[scale=.65]{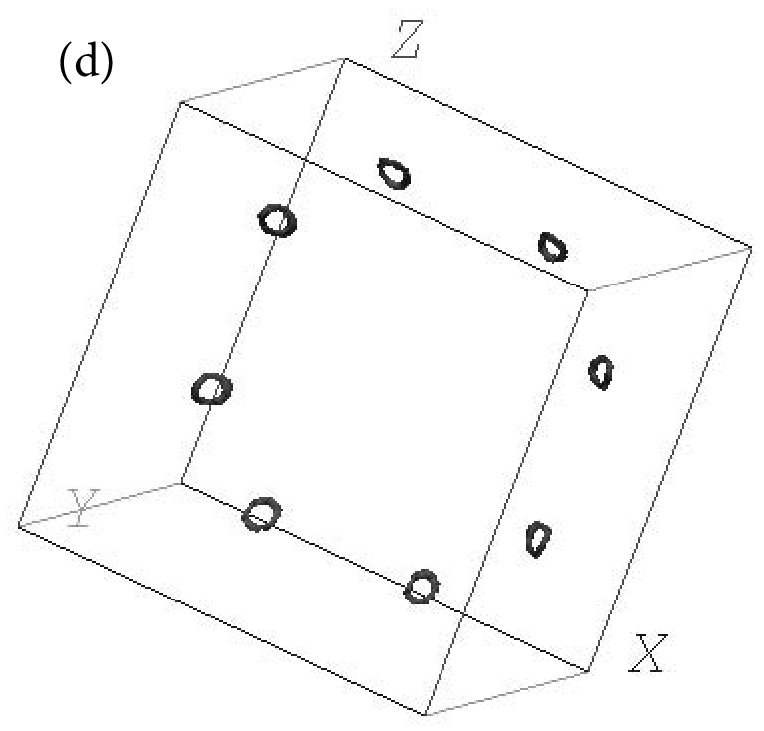}%
\hspace*{\fill}\\
\caption{Coaxial collision of two oppositely polarized circular vortices parallel to the $xz$
plane with similar radii, $R = 18$, $r = 14$ ($R/r = 1.3$). The periodicity box is not too
large ($2L = 84$, i.e. $L/R \simeq 2.33$). Consecutive times are: $0$, $8 \tau$, $14 \tau$, $18
\tau$, $\tau = 6.96$.}
\end{figure}
\begin{figure}[t]
\centering\includegraphics[scale=.65]{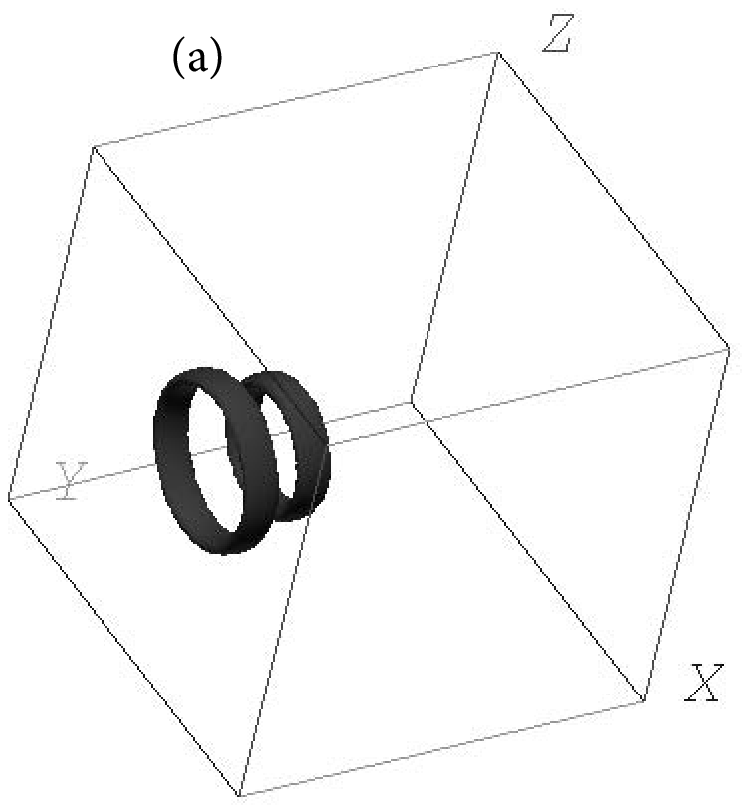}%
\quad\includegraphics[scale=.65]{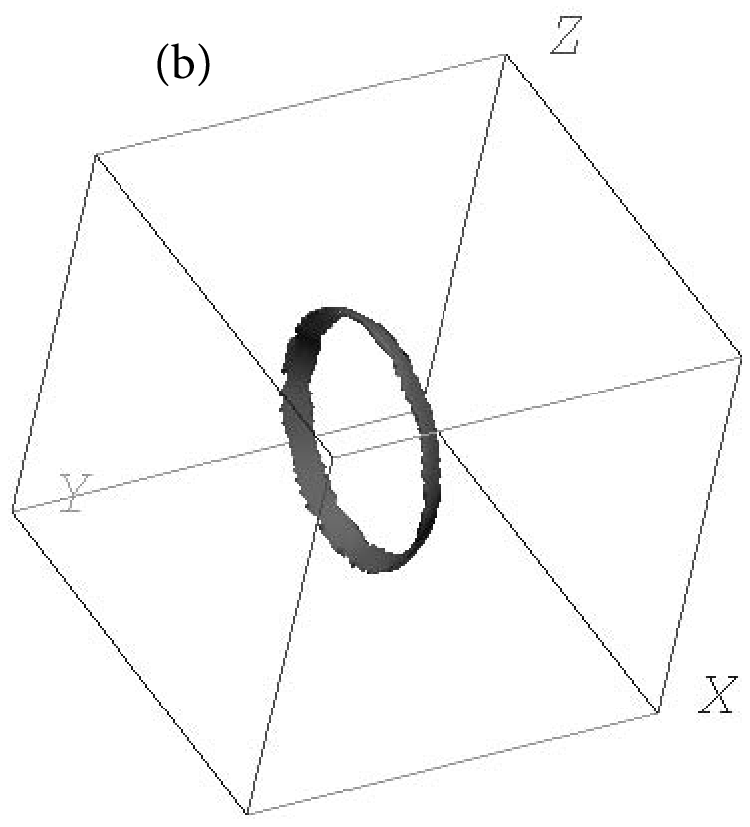}%
\quad\includegraphics[scale=.65]{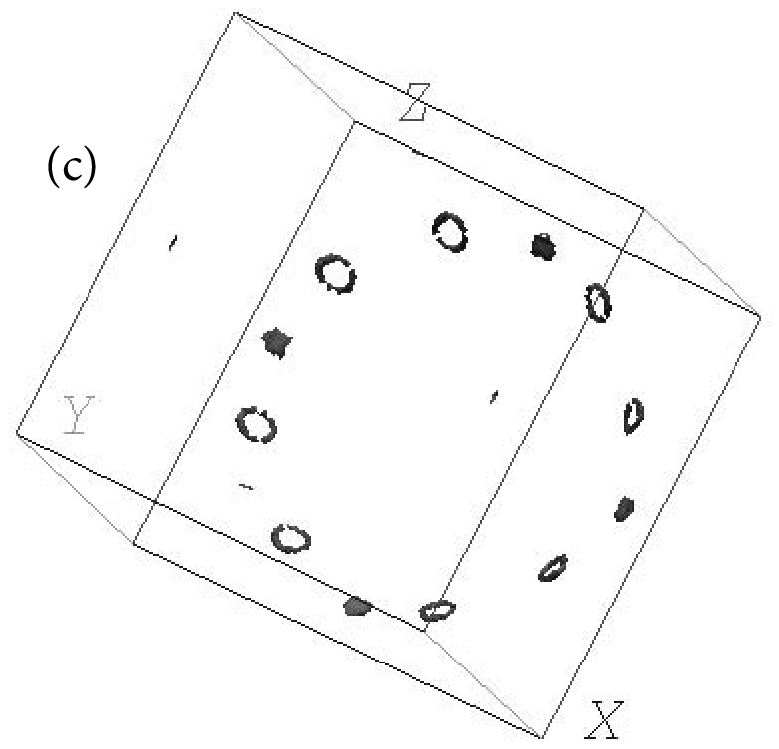}%
\hspace*{\fill}\\
\caption{Coaxial collision of two oppositely polarized circular vortices with similar radii,
identical with those shown in Fig. 6 ($R/r = 1.3$) but with the periodicity box somewhat larger
($2L = 100$, i.e. $L/R \simeq 2.78$). Consecutive times are: $0, 12 \tau, 20 \tau, \tau = 5.35$.
A ``massive'' production of smaller circular vortices continues. Note that in frame (c), the $xz$
plane is parallel to the page.}
\end{figure}

Figures 3--7 present head-on collisions of coaxially moving pairs of circular vortices
with comparable radii.
We first assume that the radius $R$ of the larger circular vortex is not too close to that of the
smaller vortex, $r$, e.g. for $R/r = 2$, see figures  3--5.

If the transverse box dimension is close enough to the size of the larger vortex, the interaction of
the colliding pair with its images in the four closest neighbouring cells predominates over the main
collision.  As a result, the structure of the periodic
vortex system changes, and four (distorted) circular vortices extending to the neighbouring cells are
formed, see figure 3, where $L/R = 1.75$. Only for more distant boundaries can we treat the
interaction of the colliding pair with neighbouring pairs
as merely a perturbation of the main collision. This perturbation can lead to the production of four
smaller circular rings after collision, if the ratio $L/R$ is not too large, see figure 4, where
$L/R = 2.625$. Otherwise, the perturbation is too weak to switch on the reconnection, and
the colliding rings pass through each other without decaying into smaller rings, see figure 5,
where $L/R = 3.45$. In all above cases ($R/r = 2$), only the interaction with four nearest cells
could be strong enough to turn on the reconnection.

The interaction with further cells can be significant if the larger ring is sufficiently close
to the smaller one, e.g. for $R/r = 1.3$, see figures 6 and 7, where respectively eight and twelve
smaller circular vortices are formed after collision.

\section{Summary}

In parts I and II the authors demonstrated how circular vortices can be created from pairs of
oppositely polarized line vortices and confirmed the $U(a)$ dependence, proving Feynman's hypothesis.
In this paper we present two alternate scenarios leading to the creation of smaller ring vortices
from larger ones.

\ackn
Use was made of the equipment of the Interdisciplinary Centre for Mathematical and
Computational Modelling (ICM), Warsaw University. Professor G. Wilk should be thanked for
exerting pressure on us to write up these results.

\appendix

\section*{Appendix}

\setcounter{section}{1}

Jones and Roberts \cite{JRG} give formulas for the wavefunction in the far field for both
the three dimensional case (circular vortex) and the two dimensional case (two counterstreaming
line vortices). One might wish to compare their 2D model with our equation (\ref{incond}) in the
far field. However, we have doubts about their derivation. They first linearize equation
(\ref{nlstr}), obtaining in the far field ($z$ is interchanged with $y$ in our notation):
\begin{equation}\label{JR4p2}
\psi_{\mathrm{i}} \sim - my[y^2 + (1 - 2 U^2)x^2]^{-1},
\end{equation}
\begin{equation}\label{JR4p3lin}
\psi_{\mathrm{r}} \sim 1 + m U [y^2 - (1 - 2 U^2)x^2][y^2 + (1 - 2 U^2)x^2]^{-2}.
\end{equation}
They then find that the leading nonlinear term $ - \case12 \psi_{\mathrm{i}}^2$ is at least of
the same order as (\ref{JR4p3lin}), $r^{-2}$, and simply tack it on:
\begin{equation}\label{JR4p3}
\psi_{\mathrm{r}} \sim 1 + m \{U [y^2 - (1 - 2 U^2)x^2] - \case12 m y^2\}[y^2 +
(1 - 2 U^2)x^2]^{-2}.
\end{equation}
This is not proper procedure. The full, nonlinear equation (\ref{nlstr}) must be solved in the far
field. Two inhomogeneous equations are obtained for $\psi_{\mathrm{i}}$ and $\psi_{\mathrm{r}}$
(the equation for $\psi_{\mathrm{i}}$ is free of $\psi_{\mathrm{r}}$). We add the good news that
their other calculation is valid in 3D, the nonlinear correction being
of higher order in $r^{-1}$ than the linear terms ($r^{-4}$ and $r^{-3}$ respectively).

\end{document}